\documentclass[prl,twocolumn,aps, superscriptaddress,showpacs]{revtex4-1}
\usepackage{amsbsy,amssymb,amsmath,bm}
\usepackage{graphicx,color,epsfig,rotate}
\usepackage{xspace,units}
\usepackage{subfigure}
\usepackage{textcomp}		%for upright \textmu
\usepackage{epstopdf}
\usepackage{ulem}

\begin{document}
\title{Quantum oscillations and the Fermi-surface topology of the Weyl semimetal NbP}

\author{J. Klotz}
\affiliation{Hochfeld-Magnetlabor Dresden (HLD-EMFL), Helmholtz-Zentrum Dresden-Rossendorf, 01328 Dresden, Germany}
\affiliation{Institut f\"ur Festk\"orperphysik, Technische Universit\"at Dresden, 01069 Dresden, Germany}

\author{Shu-Chun Wu}
\author{Chandra Shekhar}
\author{Yan Sun}
\author{Marcus Schmidt}
\author{Michael Nicklas}
\author{Michael Baenitz}
\affiliation{Max Planck Institute for Chemical Physics of Solids, 01187 Dresden, Germany}

\author{M. Uhlarz}
\affiliation{Hochfeld-Magnetlabor Dresden (HLD-EMFL), Helmholtz-Zentrum Dresden-Rossendorf, 01328 Dresden, Germany}

\author{J. Wosnitza}
\affiliation{Hochfeld-Magnetlabor Dresden (HLD-EMFL), Helmholtz-Zentrum Dresden-Rossendorf, 01328 Dresden, Germany}
\affiliation{Institut f\"ur Festk\"orperphysik, Technische Universit\"at Dresden, 01069 Dresden, Germany}

\author{Claudia Felser}
\affiliation{Max Planck Institute for Chemical Physics of Solids, 01187 Dresden, Germany}

\author{Binghai Yan}
\email{yan@cpfs.mpg.de}
\affiliation{Max Planck Institute for Chemical Physics of Solids, 01187 Dresden, Germany}
%\affiliation{School of Physical Science and Technology, ShanghaiTech University, Shanghai 200031, China}
\affiliation{Max Planck Institute for the Physics of Complex Systems, 01187 Dresden, Germany}
%\affiliation{CAS-Shanghai Science Research Center, Shanghai 201203, China}

\date{\today}

\begin{abstract}

The Weyl semimetal NbP was found to exhibit topological Fermi arcs and exotic magneto-transport properties. 
Here, we report on magnetic quantum-oscillation measurements on NbP and construct the 3D Fermi surface with the help of band-structure calculations.
We reveal a pair of spin-orbit-split electron pockets at the Fermi energy and a similar pair of hole pockets, all of which are strongly anisotropic. 
%The Fermi surface well explains the linear magnetoresistance observed in high magnetic fields by the quantum-limit scenario.
The Weyl points that are located in the $k_z \approx \pi/c$ plane are found to exist 5 meV above the Fermi energy. Therefore, we predict that the chiral anomaly effect can be realized in NbP by electron doping to drive the Fermi energy to the Weyl points.

\end{abstract}


\maketitle

% Introduction 

The recent discovery of Weyl semimetals (WSMs)~\cite{Wan2011} in transition-metal monopnictides~\cite{Weng2015,Huang2015,Xu2015TaAs,Lv2015TaAs,Yang2015TaAs} revealed an exotic topological matter.
In a WSM, the conduction and valence bands cross each other linearly at a point, called the Weyl point, 
which always comes in pairs of opposite chirality (handedness) in a lattice. 
Similar to topological insulators (TIs),  the materials surface exhibits topological states, which form Fermi arcs connecting Weyl points with opposite chirality~\cite{Nielsen1981}. 
In a system with Weyl fermions, a nonorthogonal electric field ($\textbf{E}$) and magnetic field ($\textbf{B}$) lead to an increase or decrease of carriers with finite chirality, called chiral anomaly~\cite{Adler1969,Bell1969}. 
The chiral-anomaly effect generates a topological electric current, which is characterized by a negative magnetoresistance (MR) in experiment~\cite{Nielsen1983}. 
As evidence of Weyl fermions, Fermi arcs have already been identified 
on four transition-metal monopnictides, NbP, NbAs, TaP, and TaAs by angle-resolved photoemission spectroscopy (ARPES)~\cite{Xu2015TaAs,Lv2015TaAs,Yang2015TaAs,Lv2015TaAsbulk,Liu2015NbPTaP,Xu2015NbAs,Xu2015TaP,Souma2015,Xu2015NbP} and calculations~\cite{Weng2015,Huang2015,Sun2015arc}.
Great efforts are being devoted to demonstrate the chiral anomaly~\cite{Huang2015anomaly,Zhang2015ABJ,Shekhar2015TaP,Wang:2015wm,Yang:2015vz,Du:2015TaP} and other exotic chiral magnetic effects~\cite{Shekhar2015,Moll2015,Zhang2015quantum,Yang2011QHE,Turner:2013tf,Hosur:2013eb,Vafek:2014hl,Parameswaran2014,Baum2015}.  

NbP is a compound with the lightest elements among the four WSM materials. 
By our earlier experiments, we revealed an extremely large positive MR with very high carrier mobility in NbP when $\textbf{B} \bot \textbf{E}$~\cite{Shekhar2015}.
There, the MR increased linearly in magnetic fields up to 60~T, which calls for better understanding. 
Very recently, a negative longitudinal MR ($\textbf{B}~||~\textbf{E}$) was suggested as a plausible signature of the chiral anomaly in this material~\cite{Wang:2015wm}. 
However, the interpretation of the above magneto-transport experiment remains illusive due to the lack of accurate information of the bulk Fermi surface. 
Although recent soft X-ray ARPES addressed the Weyl points in the bulk band structure (e.g.\ Refs.~\onlinecite{Xu2015TaAs,Yang2015TaAs,Lv2015TaAsbulk}), 
its energy resolution was unfortunately insufficient to locate the exact Fermi energy in meV precision.
Alternatively, the 3D Fermi surface (FS) can be accurately reconstructed through angle-resolved magnetic quantum-oscillation (QO) experiments. 
For example, the Weyl cones were already revealed in the sister compound TaP by magnetic-torque measurements~\cite{Shekhar2015TaP}.

In this work, we performed magnetic QO measurements based on the de Haas--van Alphen (dHvA)  effect 
and match the FS with band-structure calculations for NbP. 
We reveal a pair of spin-orbit-split electron pockets at the Fermi energy and a similar pair of hole pockets, 
in which one pocket nests inside the other like a Matryoshka doll for both electron and hole pockets.
The largest Fermi-surface area corresponds to a frequency of about 40~T when $B$ is along the 
crystallographic $c$ axis. This might indicate a quantum origin, which is similar to the quantum MR proposed by Abrikosov~\cite{Abrikosov1998}, for 
the linear MR observed in Ref.~[\onlinecite{Shekhar2015}] above 40~T.
Our results show, that the Weyl points that lie in the $k_z=0$ plane of the Brillouin zone (labeled W1) are 57~meV below $E_F$ and fully merged into the electron pocket,
which is inactive for the chiral anomaly.
The Weyl points that are located in the $k_z \approx \pi / c$ plane (labeled W2) are 5~meV above $E_F$, 
in which Weyl points with opposite chirality are separated by a barrier of more than 1~meV. 
In addition, we note that the deduced bulk FS and Weyl-point positions agree well with very recent ARPES experiments~\cite{Souma2015}.

NbP is a noncentrosymmetric compound in a tetragonal lattice (space group $I4_1md$, No. 109). The magnetic QO measurements are performed on high-quality NbP single crystals in fields up to $12$~T by use of a superconducting magnet in combination with a $^3$He cryostat down to temperatures of $T=0.5$~K. A 50~\textmu m thin CuBe cantilever was employed, allowing high-resolution detection also of low-frequency dHvA signals down to 0.3~T. 

\begin{figure}
	\centering
		\includegraphics[width=0.98\columnwidth]{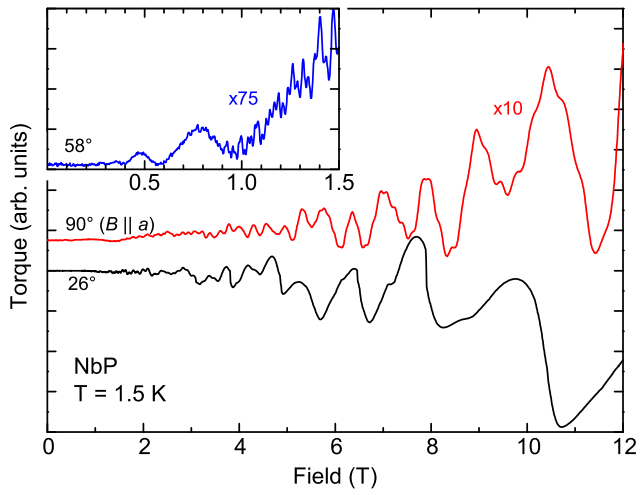}
	\caption{(color online) Torque signals for three different magnetic-field
orientations obtained at $1.5$ K. The inset shows the signal at $\Theta = 58^{\circ}$ at
low fields.}
	\label{fig:example}
\end{figure}

The crystal has been mounted in such way that the applied field, $B$, is along the $c$ axis of the crystal at $\theta$ = 0\textdegree~ and $B$ is along the $a$ axis of the crystal at $\theta$ = 90\textdegree. The angular dependences of the magnetic QO were measured at 1.5 K. 
Figure~\ref{fig:example} shows as-measured torque signals for three different field orientations.
On a smoothly varying background torque signal, clear dHvA oscillations evolve reaching very strong amplitudes (about $10$~\% of the background signal) at the highest measured fields. As shown for the data at $\Theta = 58^{\circ}$ (inset of Fig.\ \ref{fig:example}), some low-frequency quantum oscillations can be resolved already at about $0.3$~T.  That corresponds to a large magnetic length $l_B = \sqrt{h/eB} \approx 46$~nm, characterizing the extraordinary high quality of the sample and high sensitivity of the instruments. 

\begin{figure}
	\centering
		{\includegraphics[width=1\columnwidth]{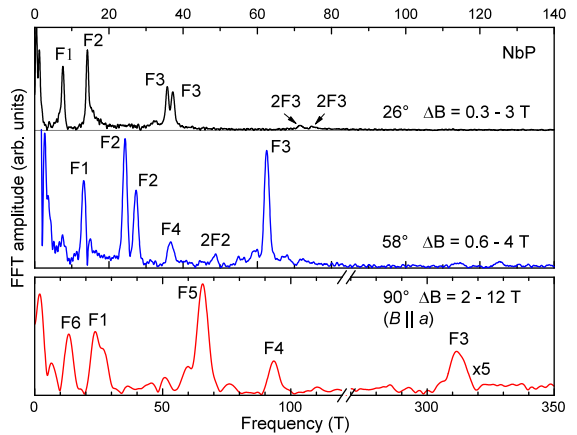}}
		\caption{(color online) Spectra of the data shown in Fig.~\ref{fig:example} after polynomial background subtraction and Fourier transformation. Note that different field ranges, $\Delta B$ were used in order to improve spectral resolution.}
		\label{fig:exampleFFT}
\end{figure}

\begin{figure}
	\centering
	{\includegraphics[width=1\columnwidth]{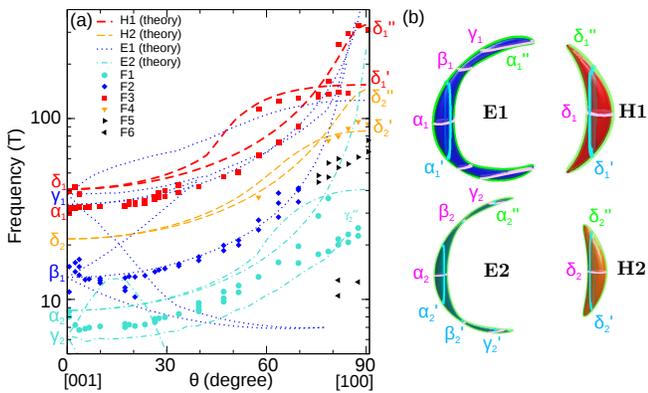}}%{ang_dep_compare.png}
	\caption{(a) Angular dependence of the measured dHvA (solid symbols) as well as the calculated quantum-oscillation frequencies (lines) for four pockets. (b) Plot of the E1, E2, H1, and H2 pockets. Pink loops show the extremal cross-sections in the (001) plane, cyan loops in the (100) plane, and green loops in the (010) plane.}
	\label{fig:dhva}
\end{figure}

Applying a fast Fourier transform (FFT) after subtracting a second-order polynomial background for our magnetic-torque measurements, we can obtain the corresponding angle-resolved dHvA-frequency spectra as shown in Fig.~\ref{fig:exampleFFT}. The observed dHvA frequencies ($F$) are related to the extremal cross-sections of the Fermi surface ($A$) via the Onsager relation \cite{shoenberg1984magnetic},
\begin{equation}
F = \hbar A/(2\pi e),
\end{equation}
 where $\hbar$ is the reduced Planck constant and $e$ is the electron charge.
For clarity, we only included those frequencies that are clearly resolved and are attributed to fundamental frequencies instead of
higher harmonics. We identify six different frequency branches, which are labeled $F1$, $F2$, $F3$, $F4$, $F5$, and $F6$. The angular dependencies of these six branches are exhibited in Fig.~\ref{fig:dhva}. For $\Theta=0^\circ$, our thermodynamic dHvA data are well consistent with earlier Shubnikov-de Haas (SdH) transport results ($F=7$, 13, and 32~T) in Ref.~[\onlinecite{Shekhar2015}], but slightly different from those ($F=14.5$, 31.2, and 64~T) in Ref.~[\onlinecite{Wang:2015wm}]. We as well observed a frequency at 64~T, but assigned that as second harmonic of our $F3$ frequency.

To fully understand the FS, we performed band-structure calculations and mapped the experimental frequencies to the calculated FS. The band-structure calculations were performed with the generalized gradient approximation (GGA) using the Vienna {\it ab-initio} simulation package ({\sc vasp})~\cite{Kresse1996}. We interpolated the bulk Fermi surface using maximally localized Wannier functions~\cite{Mostofi2008}. 

We reveal two electron pockets ($E1$ and $E2$) and two hole pockets ($H1$ and $H2$) at the Fermi energy $E_F$, all of which are strongly anisotropic as shown in Fig.~\ref{fig:dhva}(b). 
In reciprocal space, the pockets $E2$ and $H2$ lie entirely within $E1$ and $H1$, respectively. 
Both hole pockets have only one extremal orbit each for every crystallographic direction. For $H1$ ($H2$), these are labeled $\delta_1$ ($\delta_2$) for the (001) plane ($B~||~c$), $\delta_{1}'$ ($\delta_{2}'$) for the (100) plane ($B~||~a$), and $\delta_{1}''$ ($\delta_{2}''$) for the (010) plane ($B~||~a$). 
In contrast, the sickle-like shape of the electron pockets gives rise to several extremal orbits.
For the (001) plane, both electron pockets, $E1$ and $E2$, show three extremal orbits, out of which $\alpha_1$ ($\alpha_2$) results from the center, $\beta_1$ ($\beta_2$) results from the necks, and $\gamma_1$ ($\gamma_2$) results from extra humps of $E1$ ($E2$), respectively. For the (010) plane, similar to the hole pocket, only one extremal orbit, $\alpha_{1}''$ ($\alpha_{2}''$), is obtained from $E1$ ($E2$). For the (100) plane, $E1$ yields only one orbit ($\alpha_{1}'$), but three orbits ($\alpha_{2}'$, $\beta_{2}'$, and $\gamma_{2}'$) originate from $E2$. However, the cross-sections of the neck orbit of $E2$ correspond to frequencies that are always less than 1~T, which are not shown here. 
Altogether, this leads to the predicted, rather evolved, angular dependence of QO frequencies as shown in Fig.~\ref{fig:dhva}(a).

When we map the experimental frequencies to the calculated FS, the angular dependence of extremal FS cross sections agrees well with that of the measured dHvA frequencies. Our calculated FS can indicate that the $F1$, $F2$, $F3$, and $F4$ frequency branches belong to the E2, E1, H1, and H2 pocket, respectively. But some of the calculated dHvA frequencies lie very close to each other, such as the $\gamma_{1}$ orbit of $E1$ and the $\delta_1$ orbit of $H1$. Such close low-frequency frequencies are experimentally hardly separable. So we sometimes observe double-peak structures, e.g. $F3$ at $\theta$ = 26\textdegree~ in Fig.~\ref{fig:exampleFFT}, hinting at additional orbits as predicted~\cite{remark_dingle}.

Nevertheless, some of the predicted orbits of $E1$ and $E2$ were not confirmed experimentally.
A reason for that might lie in the spectral richness of our data.
Within a narrow frequency window, we resolved several fundamental
frequencies and their higher harmonics (Fig.~\ref{fig:exampleFFT}),
which render the disentanglement of different peaks challenging.
Therefore, some fundamental frequencies might be screened by other
peaks of fundamentals or harmonics.
In addition, some extremal FS cross sections might have an unfavorable
curvature factor $\left|\partial^2A/\partial k^2_{||}\right|^{-1/2}$.
This factor describes the curvature of the FS parallel to the applied field.
A strong curvature close to an extremal cross section of the FS reduces
the quantum-oscillation amplitudes, making them more difficult to resolve.

\begin{table}
	\centering
		\caption{Experimental and calculated values of the dHvA frequencies and
effective masses for different orbits and field orientations~\cite{remark_angles}. The frequency $F$ is in unit of T, and effective masses $m^{\ast}$  are in units of free electron masses ($m_e$). The W1- and W2-types of Weyl points are encompassed by the pockets E1 and H1, respectively. 
Thus, we expect that all the Fermi surfaces resolved here may exhibit a trivial Berry phase.}
		\begin{tabular*}{\columnwidth}{@{\extracolsep{\fill}}*7l@{}}%{{\extracolsep{\fill}}lllllll}
			\hline
			\hline
                      &        & \multicolumn{2}{c}{Experiment} & \multicolumn{3}{c}{Calculation} \\
                      &               & $F$ & $m^{\ast}_{\rm {exp}}$ & Pocket & $F$ & $m^{\ast}_{\rm{cal}}$ \\
			\hline
			$\alpha_{2}$ 		& $B~||~c$ &  8.7   &    0.047(9)    &  $E2$  &  9   & 0.05	\\
			$\beta_1$       & $B~||~c$ &  14.6  &    0.057(7)    &  $E1$  &   13   & 0.12	\\
			$\alpha_{1}$    & $B~||~c$ &  32.1  &    0.046(6)    &  $E1$  &   34   & 0.10	\\
			$\delta_{1}$ 		& $B~||~c$ &  39.6  &    0.052(7)    &  $H1$  &		41   & 0.07	\\
%			--              & $B~||~a$ &  12.3  &    0.066(10)   &        &        & \\
%			--          		& $B~||~a$ &  22.8  &    0.063(6)    &  		  &        & \\
%			--          		& $B~||~a$ &  25.2  &    0.07(1)     &  		  &        & \\
%			--              & $B~||~a$ &  60.5  &    0.09(1)     &        &        & \\
			$\delta_{2}'$   & $B~||~a$ &  97    &    0.33(6)     &  $H2$  &   86   & 0.20	\\
			$\delta_{1}''$  & $B~||~a$ &  324   &    0.65(10)    &  $H1$  &   331  & 0.57	\\
			\hline
			\hline
		\end{tabular*}
\label{tab:masses}
\end{table}

As mentioned, some observed fundamental frequencies can coincide with two or more orbits on the FS. 
Comparing the experimental and calculated effective masses may help to clarify the assignment.
The measured and calculated effective masses and their corresponding fundamental frequencies are summarized in Table~\ref{tab:masses}. 

We quantified the effective charge-carrier masses,
$m^{\ast}$, from the temperature dependence of the dHvA amplitudes.
According to the Lifshitz-Kosevich formula~\cite{shoenberg1984magnetic},
this temperature dependence is proportional to $X/\sinh(X)$, with
$X=\alpha m^{\ast} T/B$ and $\alpha=2\pi^2k_B m_e/(\hbar e)$.
Here, $k_B$ is the Boltzmann constant, and $m_e$ the free electron mass.
By repeating field sweeps at fixed angles close to $c$ and $a$ and different temperatures,
we determined $m^{\ast}_{\rm {exp}}$ separately for the observed dHvA orbits. 
We estimated the error by varying the field window that was used to
determine the FFT amplitudes.
For $B~||~c$, all masses are of the order of $m^{\ast}_{\rm {exp}}\approx 0.05~m_e$,
as expected for semimetals with small FS pockets.
For $B~||~a$, masses up to 0.65~$m_e$ appear~\cite{remark_angles}.
This is expected as well, since in first approximation, $m^{\ast}_{\rm {exp}}$ is proportional to the extremal
area of an orbit. This can be seen nicely for the $\delta$ orbits for which
the dHvA frequency $F$ as well as $m^{\ast}_{\rm {exp}}$ grow roughly by a factor of 10.
The effective mass can be calculated from the band structure and averaged over the FS cross section, to compare with corresponding experimental values. As listed in Table~\ref{tab:masses}, the masses from theory and experiment are qualitatively consistent.

\begin{figure}
	\centering
		{\includegraphics[width=0.98\columnwidth]{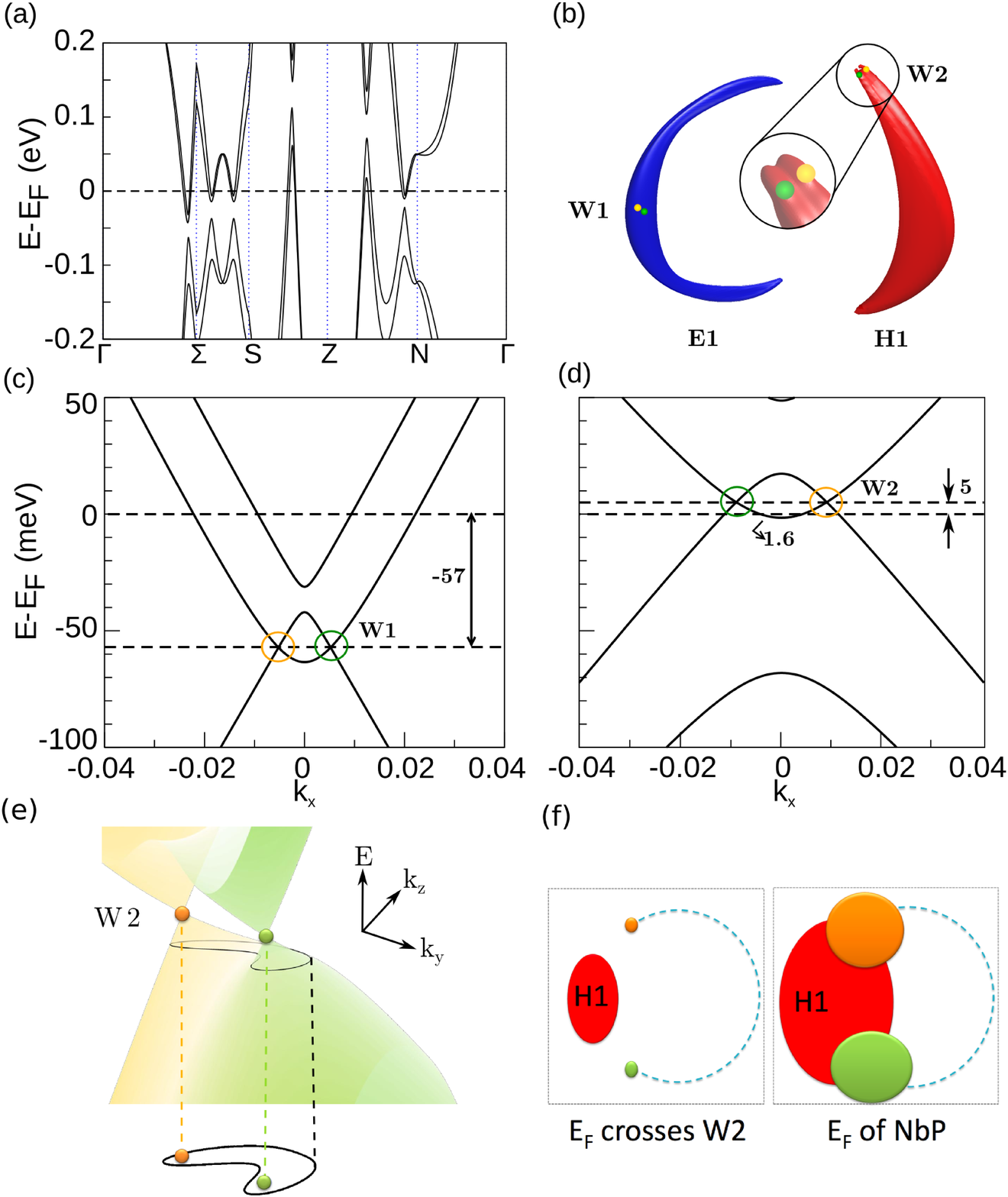}}%{NbP_bulk_band.png}
		\caption{(a) The band structure along high symmetry points. Here, the horizontal dashed lines correspond to the best-fitting Fermi energy. (c) and (d) Zoom into the $k$-space regions around the Weyl points W1 (left panel) and W2 (right panel). Both graphs are details of the band-structure reported in Ref.~\cite{Shekhar2015}. The left panel is a cut along $k_{y}=0.495\cdot2\pi/a$ and $k_{z}=0$. The right panel is a cut along $k_{y}=0.270\cdot2\pi/a$ and $k_{z}=0.571\cdot2\pi/c$. W1 and W2 were found to lie $-57$~meV below and $5$~meV above the Fermi energy, respectively. (b) The green and orange points indicate the Weyl points with opposite chirality. W1- and W2-type Weyl points can be found inside the E1 and H1 pockets, respectively. (e) The 3D band structure near the W2-type Weyl cones in the $k_y - k_z$ plane. The energy contour at the bottom corresponds to the Fermi energy measured in experiment. Two Weyl points with opposite chirality are indicated as green and yellow spheres. (f) The illustration of the Fermi arc and bulk pockets. At the current Fermi energy ($E_F$), the Fermi arc (blue dashed curve) still exists and connects two W2-type Weyl pockets that are interconnected by the H1 pocket (right panel), similar to the ideal case where $E_F$ crosses the W2-type Weyl points (left panel).}
	\label{fig:bulk_band}
\end{figure}

The best-fitting Fermi energy we got matches the charge neutral point of NbP within less than half a meV uncertainty. The calculated electron and hole carrier densities are $n_e \approx n_h = 2.5 \times10^{19}$~cm$^{-3}$, 
which compares to $n_e = 0.942\times 10^{19}$~cm$^{-3}$ and $n_h = 0.924\times 10^{19}$~cm$^{-3}$ obtained from a two-carrier model applied to the Hall measurements (we note that a single-carrier model was adopted in Ref.~\cite{Shekhar2015}). Albeit only taking two carriers into account, this model yields a good agreement between experiment and calculation. It also indicates an almost perfect electron-hole compensation leading to the extremely large MR observed previously~\cite{Shekhar2015}. The existence of carriers with tiny masses can account for the high mobility observed. Since the largest Fermi surface area ($\delta_1$) is about 40~T when $B~||~c$, the system can reach the quantum limit ($n=1$ Landau level) above about 40~T, which is consistent with the disappearance of SdH oscillations in experiment~\cite{Shekhar2015}. The W2-type Weyl points exist 5 meV above $E_F$ and are separated by a tiny barrier ($\sim$~1.6~meV) along the line connecting a pair of Weyl points, as shown in Fig.~\ref{fig:bulk_band}(d). However, the pair of W2 Weyl points merges into the hole pocket in other directions, as shown in Fig.~\ref{fig:bulk_band}(e), indicating the non-existence of independent Weyl pockets. 

Despite that the Weyl pockets are interconnected at the Fermi energy to hinder the observation of the chiral-anomaly effect, the existence of Fermi arcs, another characteristic feature of a WSM, can still exist in such a case. As illustrated in Fig.~\ref{fig:bulk_band}(f), the H1 pocket that connects a pair of W2-type Weyl pockets will not necessarily remove the Fermi arcs. We note that the Fermi-arc states can possibly survive when two Weyl pockets directly merge into each other. This is well consistent with the recent observation of Fermi-arc states at different Fermi energies in both surface calculations~\cite{Sun2015arc,Zhang2016} and ARPES experiments~\cite{Liu2015NbPTaP}.

To realize the chiral anomaly effect that is characterized by a negative longitudinal MR, it is expected that $E_F$ needs to lie as close as possible to the Weyl points. 
So we suggest that considerable electron doping is necessary for NbP to observe the chiral anomaly effect. 
To shift $E_F$ to the W2-type Weyl points,
we estimated according to our calculations that the electron and hole densities need to be doped towards $ 3.3 \times10^{19}$~cm$^{-3}$ and $ 2.2 \times10^{19}$~cm$^{-3}$, respectively.

In conclusion, our Fermi-surface study of NbP by means of dHvA measurements
and band-structure calculations gives evidence for the presence of two
electron and two hole Fermi-surface pockets. We find excellent agreement
between experiment and theory. 
The two group of Weyl nodes in NbP, W1 and W2, lie at $-57$~meV and $5$~meV
away from the Fermi energy, respectively.
This detailed knowledge of the Fermi-surface topology and the band structure
now allows to approach the Weyl points in NbP by appropriate doping and, thus,
enables the study of the interplay between dispersive and Dirac-like
quasiparticles.

We thank Dr. F. Arnold and Dr. E. Hassinger for helpful discussions.
We acknowledge support by HLD at HZDR, member of the European Magnetic
Field Laboratory (EMFL) and by the Deutsche Forschungsgemeinschaft (SFB 1143, and Project No.\ EB 518/1-1 of DFG-SPP 1666 `Topological Insulators') and by the ERC Advanced Grant [No. (291472) ``Idea Heusler''].


\begin{thebibliography}{10}
\bibitem{remark_dingle}
Note that due to the proximity of neighboring frequencies, neither a Dingle analysis for the determination of the carrier mobility nor Landau indexing for the determination of Berry phase can be performed reliably.
\bibitem{remark_angles}
Note that the actual field angle was slightly off the symmetry axes ($\sim
  1^{\circ}$) in order to obtain a larger dHvA signal.
\end{thebibliography}
\end{document}